\documentclass[12pt]{article}
\usepackage{amssymb,amsmath,epsfig}

\begin{document}
\title{\bf Effects of Electromagnetic Field on the Dynamical Instability of
Cylindrical Collapse}

\author{M. Sharif \thanks{msharif.math@pu.edu.pk} and M. Azam
\thanks{azammath@gmail.com}\\\\
Department of Mathematics, University of the Punjab,\\
Quaid-e-Azam Campus, Lahore-54590, Pakistan.}

\date{}

\maketitle
\begin{abstract}
The objective of this paper is to discuss the dynamical
instability in the context of Newtonian and post Newtonian
regimes. For this purpose, we consider non-viscous heat conducting
charged isotropic fluid as a collapsing matter with cylindrical
symmetry. Darmois junction conditions are formulated. The
perturbation scheme is applied to investigate the influence of
dissipation and electromagnetic field on the dynamical
instability. We conclude that the adiabatic index $\Gamma$ has
smaller value for such a fluid in cylindrically symmetric than
isotropic sphere.
\end{abstract}

\section{Introduction}

The stability problem of relativistic stars has been an interesting
subject for researchers. The static stellar model would be
interesting if it remains stable under fluctuations. This can be
well discussed in terms of adiabatic index $\Gamma$. The dynamical
stability \cite{1}-\cite{4} of isotropic spheres in general is given
by
$${\Gamma}\geq{\frac{4}{3}}+n\frac{M}{R},$$ where $n$ is a number
of order unity that depends on the structure of the star and $M$
and $R$ are the mass and radius of the star respectively. For
example, for white dwarfs $n=2.25$.

Different physical aspects of fluids play a key role in the
dynamical instability and evolution of self-gravitating systems. It
was shown \cite{5} that the dissipation increases the instability
range at Newtonian corrections but makes the fluid less unstable at
relativistic corrections. Chan et al. \cite{6} found that anisotropy
and radiation affect the instability range at Newtonian and post
Newtonian (pN) regimes. The same authors \cite{7} explored that the
shearing viscosity decreases the instability range of the fluid both
at Newtonian and pN regimes. Chan \cite{8} studied collapsing
radiating star with shear viscosity and concluded that shear
viscosity would increase anisotropy of pressure as well as effective
adiabatic index. Dev and Gleiser \cite{9} investigated the dynamical
instability of neutral fluid sphere with varying energy density and
showed that anisotropy enhances the stability whenever the
tangential pressure is greater than the radial pressure. Sharif and
Kausar \cite{10} discussed the dynamical instability of
expansionfree fluid at Newtonian and pN order in $f(R)$ gravity.

In a strong gravitational field, a star requires more charge to be
in stable form. The effect of electric charge in a self-gravitating
bound system has been studied by Rosseland \cite{11}. Bekenstein
\cite{12} found that charge density has a significant effect on the
system but unable to show whether the star is stable or not. Glazer
\cite{13} investigated the effect of electric charge upon the
dynamical instability of isotropic fluid. Zhang et al. \cite{14}
showed that the electric charge changes the structure of neutron
star when the charge and mass density are same. Further, the
presence of electric field increases the stability of the star
\cite{15}.

Esculpi and Aloma \cite{16} found that both anisotropy and electric
charge would boost the stability of sphere under radial adiabatic
perturbation as compared to isotropic distribution of matter.
Ernesto and Simeone \cite{17} investigated that the inclusion of
charge extends the stability regions for both bubbles and shells
around a black hole. Joshi and Malafarina \cite{18} studied the
stability of Oppenheimer, Snyder and Datt black holes under small
tangential pressure. They found that this black hole is not stable
within the collapsing cloud whose final fate of collapse is a naked
singularity rather than a black hole. In a recent paper \cite{18a},
we have discussed the effects of electromagnetic field on the
dynamical instability of a spherically symmetric expansionfree
gravitational collapse.

Some recent work \cite{19}-\cite{22} indicate great interest in
cylindrical gravitational collapse with different fluids with and
without electromagnetic field. In this paper, we use cylindrically
symmetric distribution of collapsing isotropic fluid to discuss the
dynamical instability with electromagnetic field. Darmois junction
conditions \cite{23} are used to match the interior non-rotating
cylindrically symmetric spacetime to exterior cylindrically
symmetric spacetime in the retarded time coordinates. The paper is
organized as follows. In section \textbf{2}, we formulate the
Einstein-Maxwell field equations. We discuss junction conditions in
section \textbf{3}. In section \textbf{4}, the perturbation scheme
is applied on the field as well on the dynamical equations. Section
\textbf{5} provides the discussion of the dynamical instability in
terms of non-dissipative and dissipative perturbation. We conclude
the results in the last section.

\section{Interior Fluid Distribution and the Field Equations}

We consider a collapsing cylinder bounded by a hypersurface
$\Sigma$. The line element for the interior region has the following
form
\begin{equation}\label{1}
ds^2_-=-A^2dt^{2}+B^2dr^{2}+C^2d\theta^2 +dz^2,
\end{equation}
where $A,~B$, and $C$ are functions of $t$ and $r$. To preserve the
cylindrical symmetry, we have the following constraints on the
coordinates
$$-\infty{\leq}t{\leq}\infty,\quad 0\leq{r}<\infty,
\quad -\infty<{z}<{\infty},\quad 0\leq{\theta}\leq{2\pi},$$ where we
assume comoving coordinates inside the hypersurface. The interior
coordinates are taken as
$\{\chi^{\alpha}\}=\{t,r,\theta,z\}(\alpha=0,1,2,3)$. The source in
the field equations is assumed to be locally dissipative in terms of
heat flow given as \cite{5}, \cite{5a}
\begin{equation}\label{2}
T^-_{\alpha\beta}=(\mu+p)u_{\alpha}u_{\beta}+pg_{\alpha\beta}+
q_{\alpha}u_{\beta}+q_{\beta}u_{\alpha},
\end{equation}
where $\mu$ is the energy density, $p$ is the isotropic pressure,
$u_{\alpha}$ and $q_\alpha$ are four-velocity and radial heat flux
vector of the fluid respectively satisfying
${q_\alpha}u^{\alpha}=0$. Moreover, we have
\begin{equation}\label{3}
u^{\alpha}=A^{-1}\delta^{\alpha}_{0},\quad
q^{\alpha}=q\delta^{\alpha}_{1}, \quad u^{\alpha}u_{\alpha}=-1.
\end{equation}

The Maxwell equations can be written as
\begin{eqnarray}\label{4}
F_{\alpha\beta}=\phi_{\beta,\alpha}-\phi_{\alpha,\beta},\quad
F^{\alpha\beta}_{;\beta}=4{\pi}J^{\alpha},
\end{eqnarray}
where $\phi_\alpha$ is the four potential and $J^{\alpha}$ is the
four current. The electromagnetic energy-momentum tensor is given by
\begin{eqnarray}\label{5}
E_{\alpha\beta}=\frac{1}{4\pi}\left(F^\gamma_{\alpha}F_{\beta\gamma}
-\frac{1}{4}F^{\gamma\delta}F_{\gamma\delta}g_{\alpha\beta}\right),
\end{eqnarray}
where $F_{\alpha\beta}$ is the Maxwell field tensor. Since the
charge per unit length of the cylinder is at rest with respect to
comoving coordinates, the magnetic field will be zero in this local
coordinate system \cite{22}, \cite{3a}. Thus we can write
\begin{eqnarray}\label{6}
\phi_{\alpha}=\left({\phi}(t,r),0,0,0\right),\quad
J^{\alpha}={\zeta}u^{\alpha},
\end{eqnarray}
where ${\phi}(t,r)$ and $\zeta$ is the scalar potential and charge
density. The conservation of charge requires
\begin{equation}\label{7}
s(r)=4\pi\int^r_{0}{\zeta}BCdr
\end{equation}
which is the electric charge interior to radius $r$. Using
Eq.(\ref{1}), the Maxwell equations (\ref{4}) yield
\begin{eqnarray}\label{8}
{\phi''}-\left(\frac{A'}{A}+\frac{B'}{B}-\frac{C'}{C}\right){\phi'}
&=&4\pi\zeta{AB^{2}},\\\label{9}
\dot{\phi'}-\left(\frac{\dot{A}}{A}+\frac{\dot{B}}{B}
-\frac{\dot{C}}{C}\right){\phi'}&=&0.
\end{eqnarray}
Here dot and prime represent derivatives with respect to $t$ and
$r$ respectively. Integration of Eq.(\ref{8}) with respect to $r$,
assuming $\phi'(t,0)=0$, gives
\begin{eqnarray}\label{10}
{\phi'}=\frac{AB}{C}s(r).
\end{eqnarray}
The electric field intensity is defined as
\begin{eqnarray}\label{11}
E(t,r)=\frac{s(r)}{2{\pi}C}.
\end{eqnarray}

The Einstein-Maxwell field equations,
$G^-_{\alpha\beta}=8\pi\left(T^{-}_{\alpha\beta}
+E^-_{\alpha\beta}\right)$, for the interior metric gives
\begin{eqnarray}\label{12}
4{\pi}A^{2}(2{\mu}+{\pi}E^2)&=&\left(\frac{A}{B}\right)^2
\left(\frac{B'C'}{BC}-\frac{C''}{C}\right)+\frac{\dot{B}\dot{C}}{BC},\\\label{13}
8{\pi}AB^{2}q&=&\frac{\dot{C'}}{C}-\frac{\dot{C}A'}{CA}-\frac{\dot{B}C'}{BC},\\\label{14}
4{\pi}B^{2}(2p-{\pi}E^2)&=&\left(\frac{B}{A}\right)^2
\left(\frac{\dot{A}\dot{C}}{AC}-\frac{\ddot{C}}{C}\right)+\frac{A'C'}{AC},\\\label{15}
4{\pi}(2p+{\pi}E^2)&=&\left(\frac{1}{AB}\right)
\left(\frac{A''}{B}-\frac{\ddot{B}}{A}+\frac{\dot{A}\dot{B}}{A^2}
-\frac{A'B'}{B^2}\right),\\\nonumber 4{\pi}(2p+{\pi}E^2)&=&
\frac{A''}{AB^2}-\frac{\ddot{B}}{A^2B}+\frac{\dot{A}\dot{B}}{A^3B}
-\frac{A'B'}{AB^3}+\frac{\dot{A}\dot{C}}{A^3C}-\frac{\ddot{C}}{A^2C}\\\label{16}
&-&\frac{B'C'}{B^3C}+\frac{C''}{B^2C}+\frac{A'C'}{AB^2C}-\frac{\dot{B}\dot{C}}{A^2BC}.
\end{eqnarray}
Equations (\ref{15}) and (\ref{16}) yield
\begin{eqnarray}\label{17}
\frac{1}{A^2}\left(\frac{\dot{A}\dot{C}}{AC}-\frac{\dot{B}\dot{C}}{BC}\right)
+\frac{1}{B^2}\left(\frac{A'C'}{AC}-\frac{B'C'}{BC}\right)
+\frac{1}{C}\left(\frac{C''}{B^2}-\frac{\ddot{C}}{A^2}\right)=0.
\end{eqnarray}

Thorne \cite{24} defined C-energy for the cylindrical symmetric
spacetime
\begin{equation}\label{18}
m(t,r)=\tilde{E}(t,r)=\frac{1}{8}(1-l^{-2}{\nabla}^{\alpha}\tilde{r}{\nabla}_{\alpha}\tilde{r}),
\end{equation}
The circumference radius $\rho$, specific length $l$ and areal
radius $\tilde{r}$ can be defined as
\begin{equation}\label{19}
\rho^2={\xi_{(1)a}}{\xi^a_{(1)}},\quad
l^2={\xi_{(2)a}}{\xi^a_{(2)}},\quad \tilde{r}=\rho{l},
\end{equation}
where $\xi_{(1)}=\frac{\partial}{\partial{\theta}},~
\xi_{(2)}=\frac{\partial}{\partial{z}}$ are the Killing vectors and
$\tilde{E}(t,r)$ represents the gravitational energy per unit
specific length of the cylinder. The specific energy of the cylinder
\cite{25} in the interior region with electromagnetic field can be
written as follows
\begin{equation}\label{20}
m(t,r)={E}'(t,r)=\frac{1}{8}\left[1+\left(\frac{\dot{C}}{A}\right)^2
-\left(\frac{C'}{B}\right)^2\right]+\frac{s^2}{2C}.
\end{equation}
The dynamical equations can be obtained from the
contracted Bianchi identities
$(T^{-\alpha\beta}+E^{-\alpha\beta})_{;\beta}=0$ yielding the
following two equations
\begin{eqnarray}\label{21}
\dot{\mu}+q'A+\left(2\frac{A'}{A}+\frac{B'}{B}+\frac{C'}{C}\right)qA
+({\mu}+p)\left(\frac{\dot{B}}{B}+\frac{\dot{C}}{C}\right)&=&0,\\\label{22}
p'+\dot{q}\frac{A}{B^2}+\left(3\frac{\dot{B}}{B}+\frac{\dot{C}}{C}\right)q\frac{B^2}{A}
+(\mu+p)\frac{A'}{A}-(CE'+EC')\frac{E}{C}&=&0.
\end{eqnarray}

\section{Junction Conditions}

In this section, we take a timelike $3D$ hypersurface $\Sigma$
which splits two $4D$ manifolds $V^{-}$ and $V^{+}$ as interior
and exterior regions respectively. The interior region is defined
in Eq.(\ref{1}) while the exterior region is described by
cylindrically symmetric manifold \cite{26} in the retarded time
coordinate as
\begin{equation}\label{23}
ds^2_+=-\left(-\frac{2M(\nu)}{R}+\frac{Q^2(\nu)}{R^2}\right)d\nu^2
-2d{\nu}dR+R^2(d\theta^2+{\gamma}^2dz^2),
\end{equation}
where $\gamma^2 =-\frac{\Lambda}{3}$, $\Lambda$ being the
cosmological constant and $M(\nu)$ is the total mass, $Q(\nu)$ is
the charge inside the boundary surface $\Sigma$ and $\nu$ is the
retarded time. For smooth matching of the interior and exterior
regions, Darmois conditions
\cite{23} can be stated as follows:\\\\
1. The continuity of the line elements over $\Sigma$
\begin{equation}\label{24}
\left(ds^{2}_{-}\right)_{\Sigma}=\left(ds^{2}_{+}\right)_{\Sigma}
=\left(ds^{2}\right)_{\Sigma}.
\end{equation}
2. The continuity of the extrinsic curvature over $\Sigma$
\begin{equation}\label{25}
\left[K_{ij}\right]=K^{+}_{ij}-K^{-}_{ij}=0,\quad (i,j=0,2,3).
\end{equation}
The boundary surface $\Sigma$ in terms of interior and exterior
coordinates can be defined as
\begin{eqnarray}\label{26}
f_{-}(t,r)&=&r-r_{\Sigma}=0,\\\label{27}
f_{+}(\nu,R)&=&R-R(\nu_{\Sigma})=0,
\end{eqnarray}
where $r_{\Sigma}$ is a constant. Using Eqs.(\ref{26}) and
(\ref{27}), the interior and exterior metrics take the following
form over $\Sigma$
\begin{eqnarray}\label{28}
(ds^2_{-})_{\Sigma}&=&-A^2(t,r_{\Sigma})dt^{2}+C^2(t,r_{\Sigma})d\theta^{2}+dz^2,
\end{eqnarray}
\begin{eqnarray}\nonumber
(ds^2_{+})_{\Sigma}&=&-\left(-\frac{2M({\nu})}{R(\nu)_{\Sigma}}
+\frac{Q^2({\nu})}{R(\nu)^2_{\Sigma}}+2\frac{dR(\nu)_{\Sigma}}{d\nu}\right)d\nu^2
+R^2_{\Sigma}(d\theta^2+\gamma^2{dz^2}).\\\label{29}
\end{eqnarray}

The first condition implies
\begin{eqnarray}\label{30}
\frac{dt}{d\tau}&=&A(t, r_{\Sigma})^{-1},\quad
C(t,r_{\Sigma})=R_{\Sigma}(\nu)=\frac{1}{\gamma},
\end{eqnarray}
yielding the radius of $\Sigma$ in both interior and exterior
coordinates
\begin{eqnarray}\label{31}
\left(\frac{d\nu}{d\tau}\right)^{-2}&=&\left(-\frac{2M(\nu)}{R_{\Sigma}}
+\frac{Q(\nu)^2}{R^2_{\Sigma}}+2\frac{dR_{\Sigma}}{d\nu}\right).
\end{eqnarray}
The outward unit normals to $\Sigma$ are found by using
Eqs.(\ref{26}) and (\ref{27})
\begin{eqnarray}\label{32}
n^{-}_{\alpha}&=&\left(0,B(t,r_{\Sigma}),0,0\right),\\\label{33}
n^{+}_{\alpha}&=&\left(-\frac{2M(\nu)}{R_{\Sigma}}
+\frac{Q(\nu)^2}{R^2_{\Sigma}}+2\frac{dR_{\Sigma}}{d\nu}\right)^{-\frac{1}{2}}
\left(-\frac{dR_{\Sigma}}{d\nu},1,0,0\right).
\end{eqnarray}
The non-vanishing components of extrinsic curvature in terms of
interior and exterior coordinates are
\begin{eqnarray}\label{34}
&&K^{-}_{00}=-\left[\frac{A'}{AB}\right]_{\Sigma},\quad
K^{-}_{22}=\left[\frac{CC'}{B}\right]_{\Sigma},\\\label{35}
&&K^{+}_{00}=\left[\left(\frac{d^{2}\nu}{d\tau^{2}}\right)
\left(\frac{d\nu}{d\tau}\right)^{-1}
-\left(\frac{d\nu}{d\tau}\right)\left(\frac{M}{R^{2}}
-\frac{Q^{2}}{R^{3}}\right)\right]_{\Sigma},\\\label{36}
&&K^{+}_{22}=\left[R\left(\frac{dR}{d\tau}\right)
-\left(\frac{d\nu}{d\tau}\right)\left(2M-\frac{Q^{2}}
{R}\right)\right]_{\Sigma}={\gamma}^2K^{+}_{33}.
\end{eqnarray}
Using the continuity of extrinsic curvature and Eqs.(\ref{30}) and
(\ref{31}), we get
\begin{eqnarray}\label{37}
M(\nu)=\frac{R}{2}\left[\left(\frac{\dot{R}}{A}\right)^2
-\left(\frac{R'}{B}\right)^2\right]+\frac{Q^2}{2R}.\\\label{38}
E'-M\overset{\Sigma}=\frac{1}{8}, \quad p\overset{\Sigma}=(qB).
\end{eqnarray}
where $s\overset{\Sigma}=Q$ has been used. This shows that the
difference of both masses at the boundary $\Sigma$ turns out to be
$\frac{1}{8}$ which is zero in spherical case. This is the least
satisfactory condition of the C-energy where the isotropic
pressure is equal to the radial heat-flow. It is mentioned here
that pressure at the boundary will be zero only if there is no
dissipation, i.e., $q\overset{\Sigma}=0$.

\section{The Perturbation Scheme}

A stable stationary black hole solution under perturbation
indicates the final state of the dynamical evolution of a
gravitating system. This scheme is also used to identify basic
properties of black hole solution \cite{27}. Here we perturb the
field equations, Bianchi identities and all the material functions
upto first order in $\lambda$ such that the fluid is in
hydrostatic equilibrium form. These are given by \cite{5}-\cite{7}
\begin{eqnarray}\label{39}
A(t,r)&=&A_0(r)+\lambda T(t)a(r),\\\label{40}
B(t,r)&=&B_0(r)+\lambda T(t)b(r),\\\label{41}
C(t,r)&=&rB(t,r)[1+\lambda T(t)\bar{c}(r)],\\\label{42}
E(t,r)&=&E_0(r)+\lambda T(t)e(r),\\\label{43}
\mu(t,r)&=&\mu_0(r)+\lambda \bar{\mu}(t,r),\\\label{44}
p(t,r)&=&p_0(r)+\lambda \bar{p}(t,r),\\\label{45}
q(t,r)&=&\lambda\bar{q}(t,r),\\\label{45a}
m(t,r)&=&m_0(r)+\lambda\bar{m}(t,r),
\end{eqnarray}
where $0<\lambda\ll1$. The static configuration of
Eqs.(\ref{12})-(\ref{14}) and (\ref{17}) is obtained by using
Eqs.(\ref{39})-(\ref{45}) as follows
\begin{eqnarray}\label{46}
4{\pi}\left(2\mu_0+{\pi}E^{2}_{0}\right)
&=&\frac{1}{B^2_0}\left[\left(\frac{B_0'}{B_0}\right)^2-\frac{1}{r}\frac{B'_0}{B_0}
-\frac{B''_0}{B_0}\right],\\\label{47}
4{\pi}\left(2p_0-{\pi}E^{2}_{0}\right)
&=&\frac{1}{B^2_0}\frac{A'_0}{A_0}\left(\frac{B_0'}{B_0}+\frac{1}{r}\right),\\\label{48}
0&=&\frac{1}{B^2_0}\left[\left(\frac{A_0'}{A_0}-\frac{B_0'}{B_0}\right)
\left(\frac{1}{r}+\frac{B_0'}{B_0}\right)+\left(\frac{1}{r}\frac{B_0'}{B_0}
+\frac{B''_0}{B_0}\right)\right].
\end{eqnarray}

The corresponding perturbed field equations become
\begin{eqnarray}\nonumber
8{\pi}({\bar\mu}+{\pi}{E_0}Te)&=&-\frac{T}{B_0^2}
\left[\left(\frac{b}{B_0}-\bar{c}\right)\left(\frac{B'_0}{B_0}\right)^2
+\frac{1}{B_0}\left((\bar{c}B'_0)'+\frac{(rb')'}{r}\right)\right.\\\nonumber
&+&\left.\frac{2}{r}(\bar{c}'+\frac{r\bar{c}''}{2})-\frac{B'_0}{B_0}
\left(\frac{2b'}{B_0}-\frac{\bar{c}}{r}\right)\right]
-\frac{4{\pi}T}{B^2_0}\left(\frac{3b}{B_0}+\bar{c}\right)\\\label{49}
&\times&(2\mu_0+{\pi}E^2_0),\\\label{50}
8\pi{B^2_0}\bar{q}&=&\left[\left(\frac{b}{A_0B_0}\right)'
+\left(\frac{\bar{c}}{A_0}\right)'+\left(\frac{1}{r}+\frac{B'_0}{B_0}\right)
\left(\frac{\bar{c}}{A_0}\right)\right]\dot{T},
\end{eqnarray}
\begin{eqnarray}\nonumber
8{\pi}(\bar{p}-{\pi}{E_0}Te)&=&-\frac{\ddot{T}}{A_0^2}
\left(\frac{b}{B_0}+\bar{c}\right)+\frac{T}{B_0^2}\left[\left(\frac{1}{r}
+\frac{B_0'}{B_0}\right)\left(\frac{a}{A_0}\right)'\right.\\\label{51}
&+&\left.\frac{A_0'}{A_0}\left(\frac{b}{B_0}+\bar{c}\right)'
\right]-8{\pi}\frac{Tb}{B_0}\left(2p_{0}-{\pi}E^{2}_{0}\right),\\\nonumber
0&=&-\frac{1}{A^2_0B_0}(b+\bar{c}B_0)\ddot{T}+\frac{T}{B^2_0}
\left[\frac{b}{rB_0}\left(\frac{1}{A_0}-\frac{1}{B_0}+r\left(\frac{B'_0}{B_0}\right)^2
\right.\right.\\\nonumber&+&\left.\left.\frac{B'_0}{B_0}\right)
+\frac{b'}{B_0}\left(\frac{1}{A_0}-\frac{1}{B_0}+\frac{2}{r}
-\frac{B'_0}{A_0}-\frac{B_0}{rA_0}\right)+\left(2\bar{c}'\right.\right.\\\nonumber
&-&\left.\left.\frac{aA'_0}{A^2_0}\right)
\left(\frac{1}{r}+\frac{B'_0}{B_0}\right)+\frac{a'}{A_0}\left(\frac{1}{r}
+\frac{B'_0}{B_0}\right)+\frac{b''}{B_0}
+\frac{\bar{c}}{r}\left(\frac{1}{A_0}\right.\right.\\\label{52}
&-&\left.\left.\frac{1}{B_0}+2\frac{B'_0}{B_0}
+\frac{rB''_0}{B_0}\right)\right].
\end{eqnarray}
Bianchi identities (\ref{21}) and (\ref{22}) for static and
perturbed configuration yields
\begin{eqnarray}\label{53}
&&p'_0+(\mu_0+p_0)\frac{A_0'}{A_0}-E_0\left[E_0\left(\frac{1}{r}
+\frac{B_0'}{B_0}\right)+E'_{0}\right]=0,\\\label{54}
&&\bar{\dot{\mu}}+\bar{q}'A_0+\left(2\frac{A_0'}{A_0}+\frac{B_0'}{B_0}
+\frac{1}{r}\right)\bar{q}A_0+\left(\frac{2b}{B_0}+\bar{c}\right)
(\mu_0+p_0)\dot{T}=0,\\\nonumber
&&{\bar{p}}'+\dot{\bar{q}}\frac{B^2_0}{A_0}+(\bar{p}+\bar{\mu})
\left(\frac{A'_0}{A_0}\right)+(p_0+\mu_0)\left(\frac{a}{A_0}\right)'T
-E_0\left[(E_0\bar{c})'\right.\\\label{55}&+&\left.(E_0\bar{c}+e)\left(\frac{1}{r}
+\frac{B_0'}{B_0}\right)+\frac{1}{rB_0}(rbE_0)'+e'\right]=0.
\end{eqnarray}
Using Eq.(\ref{50}) in (\ref{54}) and then integrating, it follows
\begin{eqnarray}\label{56}
\bar{\mu}=-\left(\frac{2b}{B_0}+\bar{c}\right)(\mu_0+p_0)T
-\frac{1}{8\pi}\frac{A_0}{B^2_0}\phi(r)T,
\end{eqnarray}
where
\begin{eqnarray}\nonumber
\phi(r)&=&\left\{\left(\frac{1}{r}
+\frac{2A_0'}{A_0}\right)\left[\left(\frac{b}{A_0B_0}\right)'
+\left(\frac{\bar{c}}{A_0}\right)'+\left(\frac{1}{r}+\frac{B'_0}{B_0}\right)
\left(\frac{\bar{c}}{A_0}\right)\right]\right.\\\label{57}
&+&\left.\left[\left(\frac{b}{A_0B_0}\right)'
+\left(\frac{\bar{c}}{A_0}\right)'+\left(\frac{1}{r}+\frac{B'_0}{B_0}\right)
\left(\frac{\bar{c}}{A_0}\right)\right]'\right\}.
\end{eqnarray}
Similarly, the static configuration of Eq.(\ref{20}) turn out to
be
\begin{eqnarray}\label{58}
m_0&=&\frac{1}{8}\left[1-\left(\frac{1}{r}+\frac{B'_0}{B_0}\right)^2
+16{\pi}^2E^2_0{B_0r}\right].
\end{eqnarray}
The above equation gives the total energy entrapped inside
$\Sigma$. From Eqs.(\ref{53}) and (\ref{58}), we have
\begin{eqnarray}\label{59}
\frac{A_0'}{A_0}&=&-\frac{1}{\mu_0+p_0}\left[p'_0
-E_0\left[E_0\left(\frac{1}{r}
+\frac{B_0'}{B_0}\right)+E'_{0}\right]\right],\\\label{60}
\frac{B_0'}{B_0}&=&-\frac{1}{r}+\sqrt{1-8{m_0}+16{\pi}^2E^2_0{B_0r}}.
\end{eqnarray}
The second order linear differential equation can be obtained from
Eqs.(\ref{50}) and (\ref{51}) with the help of junction condition
and the relation $p_0\overset{\Sigma}=0$
\begin{equation}\label{61}
\ddot{T}+2\beta\dot{T}-\alpha{T}\overset{\Sigma}=0,
\end{equation}
where
\begin{eqnarray}\nonumber
\alpha(r)&\overset{\Sigma}=&\left(\frac{A_{0}}{B_{0}}\right)^2
\left(\frac{b}{B_0}+\bar{c}\right)^{-1}\left[\left(\frac{1}{r}
+\frac{B_0'}{B_0}\right)\left(\frac{a}{A_0}\right)'\right.\\\label{62}
&+&\left.\frac{A_0'}{A_0}\left(\frac{b}{B_0}+\bar{c}\right)'
+8{\pi}^2E_0B_0(eb_0+bE_0)\right],\\\label{63}
\beta(r)&\overset{\Sigma}=&\frac{1}{2}\frac{A^2_0}{B_0}\left(\frac{b}{B_0}+\bar{c}\right)^{-1}
\left[\left(\frac{b}{A_0B_0}\right)'
+\left(\frac{\bar{c}}{A_0}\right)'+\left(\frac{1}{r}
+\frac{B_0'}{B_0}\right)\left(\frac{\bar{c}}{A_0}\right)\right].
\end{eqnarray}
The general solution of Eq.(\ref{61}) is given by
\begin{equation}\label{64}
T(t)=-\exp[(-\beta_{\Sigma}+\sqrt{\alpha_{\Sigma}+\beta^2_{\Sigma}})t],
\end{equation}
where for the sake of real solution and range of instability, we
assume that $\alpha_{\Sigma}>0$ and $\beta_{\Sigma}<0$. We see that
the system starts collapsing at $t=-\infty$ with $T(-\infty)=0$ in
static position and it continues to collapse with the increase of
$t$.

\section{Dynamical Instability}

Here we discuss the dynamical instability at Newtonian
and post-Newtonian regimes. We relate $\Gamma$ in terms of
$\bar{p}$ and $\bar{\mu}$ as the ratio of specific heats
\begin{eqnarray}\label{65}
\bar{p}=\Gamma\frac{p_0}{\mu_0+p_0}\bar{\mu},
\end{eqnarray}
$\Gamma$ is taken to be constant throughout the dynamical
instability. Using Eq.(\ref{56}) in the above equation, we obtain
\begin{eqnarray}\label{66}
\bar{p}=-{p_0}\Gamma\left(\frac{2b}{B_0}+\bar{c}\right)T-
\frac{1}{8\pi}\frac{A_0}{B^2_0}\Gamma\frac{p_0}{\mu_0+p_0}\phi(r)T.
\end{eqnarray}
Also, from Eq.(\ref{51}), it follows that
\begin{eqnarray}\nonumber
\left(\frac{a}{A_0}\right)'&=&\left(\frac{1}{r}+\frac{B'_0}{B_0}\right)^{-1}
\left[-8\pi\left(\frac{2b}{B_0}+\bar{c}\right)\Gamma{p_0}B^2_0
-{\Gamma}\frac{p_0}{\mu_0+p_0}A_0\phi(r)T\right.\\\nonumber
&+&\left.\frac{B^2_0}{A_0}\left(\frac{b}{B_0}+\bar{c}\right)\frac{\ddot{T}}{T}
-\frac{A'_0}{A_0}\left(\frac{b}{B_0}+\bar{c}\right)'-8\pi^2E_0B_0(eB_0+bE_0)
\right.\\\label{67} &+&\left. 16{\pi}bB_0p_0\right].
\end{eqnarray}
Making use of Eqs.(\ref{50}), (\ref{56}), (\ref{66}) and
(\ref{67}) in (\ref{55}), we get the collapse equation
\begin{eqnarray}\nonumber
&&\left[-{p_0}\Gamma\left(\frac{2b}{B_0}+\bar{c}\right)-
\frac{1}{8\pi}\frac{A_0}{B^2_0}\Gamma\frac{p_0}{\mu_0+p_0}\phi(r)\right]'
-\left[\left(\frac{2b}{B_0}+\bar{c}\right)(\mu_0+p_0\right.\\\nonumber
&+&\left.\Gamma{p_0})-\frac{1}{8\pi}\frac{A_0}{B^2_0}
(\Gamma\frac{p_0}{\mu_0+p_0}+1)\phi(r)\right]\frac{A'_0}{A_0}
+(\mu_0+p_0)\left(\frac{1}{r}+\frac{B'_0}{B_0}\right)^{-1}\\\nonumber
&\times&\left[-8\pi\left(\frac{2b}{B_0}+\bar{c}\right)\Gamma{p_0}B^2_0
-{\Gamma}\frac{p_0}{\mu_0+p_0}A_0\phi(r)+\frac{B^2_0}{A_0}\left(\frac{b}{B_0}
+\bar{c}\right)\frac{\ddot{T}}{T}\right.\\\nonumber
&-&\left.\frac{A'_0}{A_0}\left(\frac{b}{B_0}+\bar{c}\right)'-8\pi^2E_0B_0(eB_0+bE_0)
+16{\pi}bB_0p_0\right]-E_0\left[e'\right.
\\\label{68}&+&\left.(E_0\bar{c})'+(E_0\bar{c}+e)\left(\frac{1}{r}
+\frac{B_0'}{B_0}\right)+\frac{1}{rB_0}(rbE_0)'\right]
+\frac{1}{4\pi}\frac{B_0}{A^3_0}{\beta}\frac{\ddot{T}}{T}=0.
\end{eqnarray}
Now we discuss the dynamical instability at Newtonian and pN
regimes in the following cases of vanishing and non-vanishing heat
flow.

\subsection{Non-Dissipative Perturbation}

In this case, we assume that the fluid is non-dissipative, i.e.,
$q=0$. As a result, Eq.(\ref{50}) can be written as
\begin{eqnarray}\label{69}
\left(\frac{b}{A_0B_0}\right)'
+\left(\frac{\bar{c}}{A_0}\right)'+\left(\frac{1}{r}+\frac{B'_0}{B_0}\right)
\left(\frac{\bar{c}}{A_0}\right)=0.
\end{eqnarray}
This is trivially satisfied if both the terms are taken to be
identically zero, i.e.,
\begin{eqnarray}\label{70}
\left(\frac{b}{A_0B_0}\right)'=0,\quad
\left(\frac{\bar{c}}{A_0}\right)'+\left(\frac{1}{r}+\frac{B'_0}{B_0}\right)
\left(\frac{\bar{c}}{A_0}\right)=0.
\end{eqnarray}
Integration of the first term gives
\begin{eqnarray}\label{71}
b=A_0B_0
\end{eqnarray}
while the remaining two terms yield
\begin{eqnarray}\label{72}
\bar{c}'=-\left(\frac{p'_0}{\mu_0+p_0}-\frac{E^2_0+E_0E'_0+8{\pi}^2rE^4_0}{\mu_0+p_0}\right)
\bar{c}-\bar{c}\sqrt{1-8{m_0}+16{\pi}^2E^2_0{B_0r}}.
\end{eqnarray}
Since there is no dissipation, thus Eqs.(\ref{57}) and (\ref{63})
gives $\phi(r)=0$ and $\beta(r)=0$ respectively. Now using these
results and Eqs.(\ref{71}) and (\ref{72}) in the collapse equation
(\ref{68}), we obtain
\begin{eqnarray}\nonumber
&&-(2A_0+\bar{c})\Gamma{p'_0}-\Gamma{p_0}\left[(2A_0+\bar{c})\left(-\frac{p'_0}{\mu_0+p_0}
+\frac{E^2_0+E_0E'_0+8{\pi}^2rE^4_0}{\mu_0+p_0}\right)\right.\\\nonumber
&-&\left.\bar{c}\sqrt{1-8{m_0}+16{\pi}^2E^2_0{B_0r}}\right]-(2A_0+\bar{c})
(\mu_0+p_0+\Gamma{p_0})\left(-\frac{p'_0}{\mu_0+p_0}\right.\\\nonumber
&+&\left.\frac{E^2_0+E_0E'_0+8{\pi}^2rE^4_0}{\mu_0+p_0}\right)
+(\mu_0+p_0)\left(\sqrt{1-8{m_0}+16{\pi}^2E^2_0{B_0r}}\right)^{-1}
\\\nonumber&\times&\left\{-8\pi(2A_0+\bar{c})\Gamma{p_0}B^2_0-8{\pi}^2E_0(e+A_0E_0)B^2_0
+16\pi{p_0}B^2_0+\frac{B^2_0}{A_0}(A_0+\bar{c})\right.\\\nonumber&\times&
\left.\alpha_{\Sigma}-A_0\left(-\frac{p'_0}{\mu_0+p_0}+\frac{E^2_0+E_0E'_0
+8{\pi}^2rE^4_0}{\mu_0+p_0}\right.\right)^2
-\left(-\frac{p'_0}{\mu_0+p_0}\right.\\\nonumber&+&\left.
\frac{E^2_0+E_0E'_0+8{\pi}^2rE^4_0}{\mu_0+p_0}\right)\left[\left(-\frac{p'_0}{\mu_0+p_0}
+\frac{E^2_0+E_0E'_0+8{\pi}^2rE^4_0}{\mu_0+p_0}\right)
\bar{c}\right.\\\nonumber&-&\left.\left.\bar{c}\sqrt{1-8{m_0}+16{\pi}^2E^2_0{B_0r}}\right]\right\}
-E_0\left[e'+E'_0\bar{c}+A_0E'_0+\frac{A_0E_0}{r}+(E_0\bar{c}\right.
\end{eqnarray}
\begin{eqnarray}\nonumber
&+&e)\left.\left(\sqrt{1-8{m_0}+16{\pi}^2E^2_0{B_0r}}\right)-
\bar{c}E_0\sqrt{1-8{m_0}+16{\pi}^2E^2_0{B_0r}}\right.\\\nonumber
&+&\left.E_0\bar{c}\left(-\frac{p'_0}{\mu_0+p_0}
+\frac{E^2_0+E_0E'_0+8{\pi}^2rE^4_0}{\mu_0+p_0}\right)+E_0A_0\left(-\frac{p'_0}{\mu_0+p_0}
\right.\right.\\\label{73}
&+&\left.\left.\frac{E^2_0+E_0E'_0+8{\pi}^2rE^4_0}{\mu_0+p_0}-\frac{1}{r}+
\sqrt{1-8{m_0}+16{\pi}^2E^2_0{B_0r}}\right)\right]=0.
\end{eqnarray}

\subsubsection*{Newtonian limit}

For the instability condition at Newtonian limit, we take $A_0=1$,
$B_0=1$, ${\mu_0}\gg {p_0}$ and neglecting the relativistic
effects. Equation (\ref{73}) yields
\begin{eqnarray}\label{74}
-2p'_0\Gamma+2p'_0-3(E^2_0+8{\pi}^2rE^4_0+E_0E'_0)+(1-8{\pi}^2rE^2_0)
\alpha_{\Sigma}{\mu_0}=0,
\end{eqnarray}
Thus the instability condition for the isotropic fluid in the
cylindrically symmetric spacetime is given by
\begin{eqnarray}\label{75}
\Gamma<1+\left[\frac{3(E^2_0+8{\pi}^2rE^4_0+E_0E'_0)}{2\mid{p'_0}\mid}\right],
\end{eqnarray}
where we assume $p'_0<0$. This equation shows that the
electromagnetic field increases the instability of the collapsing
fluid.

\subsubsection*{Post-Newtonian limit}

In this limit, we use $A_0=1-\frac{m_0}{r},~B_0=1-\frac{m_0}{r}$
and the relativistic effects upto order $\frac{m_0}{r}$. Thus we
have
\begin{eqnarray}\nonumber
&&-2p'_0\Gamma+2p'_0-3(E^2_0+8{\pi}^2rE^4_0+E_0E'_0)+(1-8{\pi}^2rE^2_0)
\left[16\pi{p_0\mu_0}\right.\\\nonumber
&\times&(1-\Gamma)-8\pi^2{\mu_0}E_0(e+E_0)\left.+(1+\bar{c})\alpha_{\Sigma}{\mu_0}\right]
-E_0\left[e'+e(1+8{\pi}^2rE^2_0)\right.\\\label{76}&+&\left.E'_0{\bar{c}}+
(1+\bar{c})(E^2_0+8{\pi}^2rE^4_0+E_0E'_0)\frac{E_0}{\mu_0}\right]=0.
\end{eqnarray}
Since the system starts collapsing as $t{\rightarrow}-\infty$,
i.e., $\dot{T}<0$, it follows from Eq.(\ref{50}) that $\bar{c}>0$.
Hence the instability condition at pN limit becomes
\begin{eqnarray}\nonumber
&&\Gamma<1+\frac{1}{2\mid{p'_0}\mid}\left[3(E^2_0+8{\pi}^2rE^4_0+E_0E'_0)
+8\pi^2{\mu_0}eE^2_0+E_0\left(e'\right.\right.\\\label{77}
&+&\left.\left.e(1+8\pi^2r{E^2_0})+E'_0{\bar{c}}\right)\right]-\frac{1}{2\mid{p'_0}\mid}
\left(16\pi{p_0}{\mu_0}+(1+\bar{c}) \alpha_{\Sigma}{\mu_0}\right).
\end{eqnarray}
This shows that the instability range is increased by the first
three terms in the square brackets and is decreased by the last
term. It is mentioned here that the first term in the square
brackets is the Newtonian term while the remaining terms are the
relativistic corrections.

\subsection{Dissipative Perturbation}

When $q\neq0$, the perturbed solution corresponding to
Eq.(\ref{71}) becomes
\begin{eqnarray}\label{78}
b(r)=A_0B_0\left[1+\eta{f(r)}\right],
\end{eqnarray}
where $\eta>0$ is of the order $\frac{m_0}{r}$. Using the above
solution in Eqs.(\ref{57}) and (\ref{63}), $\phi(r)$ and
$\beta(r)$ turn out to be of order $\frac{m_0}{r}$.

\subsubsection*{Newtonian limit}

Applying the same approach as defined in the non-dissipative case
and using the values $A_0=1,~B_0=1,~{\mu_0}\gg {p_0}$,
Eq.(\ref{68}) at Newtonian limit reduces to
\begin{eqnarray}\nonumber
&&-2(1+{\eta{f}})p'_0\Gamma+2(1+{\eta{f}})p'_0-3(1+{\eta{f}})(E^2_0+E_0E'_0+8{\pi}^2rE^4_0)
\\\label{79}&+&(1-8\pi^2rE^2_0)(1+{\eta{f}}){\mu_0}\alpha_{\Sigma}
+\frac{\eta{f'}}{4\pi}\alpha_{\Sigma}=0.
\end{eqnarray}
In general, the heat flow $q>0$ and $\dot{T}<0$ for a collapsing
fluid, implying $\bar{c}<0$ and $f'<0$ from Eq.(\ref{50}).
Consequently the instability condition becomes
\begin{eqnarray}\label{80}
\Gamma<1+\left[\frac{3(E^2_0+E_0E'_0+8{\pi}^2rE^4_0)}{2\mid{p'_0}\mid}
+\frac{\eta{\mid{f'}\mid}\alpha_{\Sigma}}{{8{\pi}\mid{p'_0}\mid}(1+{\eta{\mid{f}\mid}})}\right].
\end{eqnarray}
We note that the electromagnetic field as well as dissipation
enhance the instability range, which supports the result for
spherical case \cite{5}.

\subsubsection*{Post-Newtonian limit}

Here we take $A_0=1-\frac{m_0}{r},\quad B_0=1-\frac{m_0}{r}$ and
consider the relativistic correction terms upto order
$\frac{m_0}{r}$. The collapse equation at pN limit yields
\begin{eqnarray}\nonumber
&&-(2+2{\eta{f}}+\bar{c})p'_0\Gamma+(2+2{\eta{f}}+\bar{c})p'_0
-(3+3{\eta{f}}+\bar{c})(E^2_0+E_0E'_0\\\nonumber&+&8{\pi}^2rE^4_0)+
\frac{1}{4\pi}\eta{f'}\alpha_{\Sigma}-(1-8\pi^2rE^2_0)\left[8\pi^2{\mu_0}E_0e
+8\pi(2+2{\eta{f}}+\bar{c})\right.\\\nonumber&\times&\left.p_0{\mu_0}\Gamma-(1+{\eta{f}}+\bar{c})
\alpha_{\Sigma}{\mu_0}-8\pi(1+{\eta{f}})(2p_0-\pi{E_0})\mu_0+({\eta{f}'}+\bar{c}')
\right.\\\nonumber&\times&\left.(-p'_0+E^2_0+E_0E'_0+8{\pi}^2rE^4_0)\right]
-E_0\left[e'+(1+8\pi^2rE^2_0)(e+E_0\bar{c})\right.
\end{eqnarray}
\begin{eqnarray}\label{81}
\left.+E_0\bar{c}'+E'_0\bar{c}+E_0\left(\eta{f}'+\frac{(1+\eta{f})}{\mu_0}
(E^2_0+E_0E'_0+8{\pi}^2rE^4_0)\right)\right]=0.
\end{eqnarray}
Using the same arguments as given in Newtonian limit, we have the
instability condition
\begin{eqnarray}\nonumber
&&\Gamma<1+\left\{\frac{3(E^2_0+E_0E'_0+8{\pi}^2rE^4_0)}{2\mid{p'_0}\mid}
+\frac{1}{(1+\eta{\mid{f}\mid})}\frac{\eta{\mid{f'}\mid}}{8\pi{\mid{p'_0}\mid}}\alpha_{\Sigma}
\right.\\\nonumber&+&\left.
\frac{4\pi^2{\mu_0}E_0e}{(1+{\eta{\mid{f}\mid}}){\mid{p'_0}\mid}}
+\frac{E_0}{(1+\eta{\mid{f}\mid})\mid{p'_0}\mid}
\left[e'+\bar{c}\alpha_{\Sigma}{\mu_0}+\eta{\mid{f'}\mid}E_0'\right.\right.\\\nonumber&+&
\left.\left.e(1+8\pi^2rE^2_0)\right]\right\}
-E_0\frac{(E_0\bar{c})'+E_0\bar{c}}{(1+\eta{\mid{f}\mid})\mid{p'_0}\mid}
-\frac{\alpha_{\Sigma}{\mu_0}}{{\mid{p'_0}\mid}}
\\\label{82}&-&\frac{4\pi(2p_0+{\pi}E_0){\mu_0}}{{\mid{p'_0}\mid}}
-\eta{\mid{f'}\mid}\left(1+\frac{E^2_0+E_0E'_0+8{\pi}^2rE^4_0}{{\mid{p'_0}\mid}}\right).
\end{eqnarray}
This implies that the range of instability is increased by
Newtonian term and relativistic terms in the square brackets while
it is diminished by the last four terms.

\section{Summary}

This paper has been addressed to discuss the dynamical instability
of cylindrically symmetric non-viscous heat conducting isotropic
fluid with electromagnetic field. For spherically symmetric
spacetime \cite{1}, it has been shown that when a star is compressed
or expanded adiabatically, its dynamical instability depends on the
numerical value of the adiabatic index, i.e., $\frac{4}{3}$. If
$\Gamma>\frac{4}{3}$, the pressure in the star is strong enough than
the weight of the layers which makes the star stable. On the other
hand, If $\Gamma<\frac{4}{3}$, the weight increases very fast than
the pressure and the star collapses resulting a dynamical
instability.

We have found that for the cylindrically symmetric case, the
dynamical instability depends on the critical value $1$. When
$\Gamma>1,$ the stability of the system is obtained while
$\Gamma<1$ yields a dynamical instability. Further, we have
studied the role played by the electromagnetic field and
dissipation in the early stages of dynamical instability. It is
interesting to mention here that different ranges of instability
may lead to different patterns of evolution of stars. We note from
Eqs.(\ref{75}) and (\ref{80}) that electromagnetic field as well
as dissipation make the system more unstable at Newtonian regime.
This supports the fact that different physical aspects of the
fluid has a great relevance in the evolution as well as structure
formation of self-gravitating objects. A similar remark applies
for Eqs.(\ref{77}) and (\ref{82}) at pN regime. It is mentioned
here that only relativistic effects are taken into account at pN
regime. We have found that the results of dissipative case reduce
to non-dissipative if heat flow vanishes.

\end{document}